\documentclass[pra,superscriptaddress,twocolumn,showpacs,preprintnumbers,amsmath,amssymb,fixfloat]{revtex4-2}
\usepackage{graphicx}% Include figure files
\usepackage{amssymb}
\usepackage{dcolumn}% Align table columns on decimal point
\usepackage{bm}% bold math 0.011{\rm ~eV}
\usepackage{bbm}
\usepackage{float}
\usepackage{siunitx}
\usepackage{amsmath}
\usepackage{multirow}
\usepackage{booktabs}
\usepackage{array}

\usepackage{amsmath, amssymb}
\usepackage{bbold}
\usepackage{makecell}
\usepackage{color} 
\usepackage{multirow}
\usepackage{mathtools}
\def\block(#1,#2)#3{\multicolumn{#2}{c}{\multirow{#1}{*}{$ #3 $}}}

\definecolor{darkblue}{rgb}{0,0,.5} 

\begin{document}
\title{Random-time quantum measurements}
%from Gaussian oversampling to Poission undersampling
\author{Markus Sifft}
\address{Ruhr University Bochum, Faculty of Physics and Astronomy, Experimental Physics VI (AG), Germany}
\author{Daniel H\"agele}
\address{Ruhr University Bochum, Faculty of Physics and Astronomy, Experimental Physics VI (AG), Germany}
\date{\today}
\begin{abstract}
The analysis of a continuous measurement record $z(t)$ poses a fundamental challenge in quantum measurement theory. Different approaches have been used in the past as records can, e.g., exhibit predominantly Gaussian noise, telegraph noise, or clicks at random times.
The last case may appear as photon clicks in an optical spin noise measurement at very low probe laser power. Here we show that such random-time quantum measurements can similarly to the first two cases be analyzed in terms of higher-order temporal correlations of the detector output $z(t)$ and be related to the Liouvillian of the measured quantum system. Our analysis in terms of up to fourth-order spectra (quantum polyspectra) shows that this new type of spectra reveals the same valuable information as previously studied higher-order spectra in case of usual continuous quantum measurements. Surprisingly, broad-band system dynamics is revealed even for deliberately low average measurement rates. Many applications are envisioned in high-resolution spectroscopy, single-photon microscopy, circuit quantum electrodynamics, quantum sensing, and quantum measurements in general.    
\end{abstract}

 \maketitle
\section{Introduction}
\label{sec:introduction}
Continuous quantum measurements comprise a large class of experiments that includes experiments of nano-electronics \cite{ubbelohdeNATCOMM2012}, optical spin noise spectroscopy \cite{clerkRMP2010}, circuit quantum electrodynamics \cite{blaisNATUREPHYS2020}, quantum sensing \cite{degenRMP2017}, and many other fields of physics. A clear understanding and modeling of such experiments and their measurement record is a prerequisite for uncovering information about a measured quantum system. A theoretical treatment of measurement records is challenging as they can present themselves in fundamentally different ways including (i) mainly Gaussian noise, (ii) switching behavior (telegraph noise), or (iii) occurrences of peaks at random times. 
Case (i) is well-known from spin noise spectroscopy where the shot noise of the probe laser dominates the system-related signal \cite{oestreichPRL2005, mullerPHYSICAE2010, aleksandrovJP2011}. Spin noise experiments have been modeled by semiclassical Langevin theories \cite{glazovPRB2012}, path-integral approaches \cite{sinitsynRPP2016}, and also via the stochastic master equation \cite{hagelePRB2018}. Case (ii) is often observed in nano-electronics where, e.g., charge fluctuations lead to switching behavior between different levels of the detector signal \cite{kurzmannPRL2019, kamblyPRB2011,stegmannPRB2015}. The full counting statistics of switching events is the most prominent theory for evaluating such measurement traces \cite{LevitovMathPhy1996,flindtPNAS2009,cookPRA1981}. Case (iii) may occur in experiments where single photons are detected by a photo-multiplier or an avalanche photodiode. Such a situation naturally occurs in spin noise experiments at very low probe-light levels. The random appearance of clicks obeys (similar to radioactive decay) a Poisson distribution. Effectively, a probe photon  interacts at random, almost discrete times with the quantum system. The interaction with the system can affect the polarization state of the photon and lead to entanglement with the system. Consequently, the probe photon polarization contains some information on the system that may be revealed in a measurement of the probe photon.
 Such (indirect) random-time measurements of a quantum system constitute an important class of experiments where a general theory has been lacking so far. We will show that higher-order correlations of the system dynamics can be recovered from the detector output even if the average sampling rate is much below the typical frequencies set by the system dynamics.
 
Our theory is based on the stochastic master equation (SME) for modeling the detector output $z(t)$
\cite{jacobsCP2006,barchielliNC1982,barchielliBOOK2009,belavkinConf1987, diosiPLA1988,gagenPRA1993,korotkovPRB1999,korotkovPRB2001,goanPRB2001}
and on higher-order spectra of $z(t)$ \cite{hagelePRB2018,hagelePRB2020E}. In comparison with usual second-order spectra, so-called quantum polyspectra allow for a more elaborate characterization of quantum systems. We recently used quantum polyspectra to unify the cases (i) and (ii) in a single theory \cite{sifftPRR2021}. As a result, we were able to extract valuable parameters of a measured quantum system by comparing experimental and theoretical quantum polyspectra.  We show here that case (iii) can also be treated within the quantum polyspectra approach by including the detector physics into the master equation. This establishes a very general framework for modeling and evaluating random-time quantum measurements. Our theory goes beyond earlier approaches of treating randomly measured quantum dynamics. 
Gross \textit{et al.} derived a new effective SME for the random-time regime but did not attempt to develop a theory for the detector output $z(t)$ and its polyspectra \cite{Gross2018}.
Ruskov \textit{et al.} had treated only a two-level quantum system and calculated the second-order spectrum of $z(t)$ establishing the equivalence with 
 case (i) -  a continuously measured system \cite{ruskovPhysRevB2003}. Similarly, Li \textit{et al.} simulated a frequency-dependent higher-order cumulant of a randomly sampled two-level system but neither gave analytic expressions for spectra nor established a connection with the corresponding continuous measurement \cite{liNJP2013}. 
 
In the following, we shortly review the quantum polyspectra approach to continuous quantum measurements. We then explicitly include the 
interaction of the measured system with a randomly arriving probe system and the subsequent detection of the probe. Next, we calculate second-order spectra for a random-in-time measured single electron spin precessing in a magnetic field  [case (iii)]. At increasing average sampling rates, we find a transition to the quantum Zeno regime. Eventually, we calculate second-, third-, and fourth-order spectra of a nontrivial coupled spin system demonstrating the appearance of  additional and common features compared to spectra of the usual continuous measurement regime \mbox{[case (i)]}.

\section{Quantum polyspectra}
The raw experimental detector output $z(t)$ is the central quantity in the quantum polyspectra approach to continuous quantum measurements. The approach relates correlations in the measured detector output to theoretical predictions about these correlations from a quantum theory \cite{hagelePRB2018,sifftPRR2021}.  Comparing theoretical and experimental quantum polyspectra 
gives the exciting opportunity to access parameters of the measured system.
The measured detector output $z(t)$ is intrinsically noisy and must be evaluated with suitable statistical methods. 
Brillinger’s polyspectra $S_z^{(n)}$ from classical signal processing
\begin{eqnarray}
2\pi \delta(\omega_1+...+\omega_n)&S_z^{(n)}&(\omega_1,...,\omega_{n-1})\nonumber\\ 
&=& C_n(z(\omega_1),..., z(\omega_n)) \label{eq:defPolyspectra}
\end{eqnarray} 
are an uncompromising approach to characterizing a stochastic stationary process $z(t)$ \cite{Brillinger1965}.  
They generalize the usual second-order power spectrum of $z(t)$ to higher orders. They are defined via $n$th-order cumulants $C_n$  of the Fourier transform, $z(\omega) = \int z(t) e^{i \omega t} \,dt$, of the detector output. Cumulants of increasing order can be expressed in terms of products of moments starting with $C_2(x,y) = \langle x y \rangle -  \langle x  \rangle \langle  y \rangle$ and $C_3(x,y,z) = \langle (x -   \langle x  \rangle)( y -   \langle y  \rangle)( z -   \langle z  \rangle) \rangle$ getting more intricate for $n \ge 4$ \cite{gardinerBOOK2009,hagelePRB2018}. 
The second-order spectrum is identical with the power spectrum $S_z^{(2)} \propto \langle z(\omega) z^*(\omega) \rangle + ...$, where $\langle … \rangle$ relates to an average over  infinitely many possible outcomes of $z(t)$ and its Fourier transforms $z(\omega) $. In case of a real-world experiment, $S_z^{(2)}$ needs to be estimated from a finite amount of data (see  \cite{sifftPRR2021}, App. B).
$S_z^{(2)}$ exhibits, e.g., peaks at the precession frequencies of an electron spin when $z$ relates to its orientation perpendicular to the magnetic field.
The bispectrum
 $S_z^{(3)}(\omega_1,\omega_2)$ is related to $\langle z(\omega_1) z(\omega_2) z^*(\omega_1+\omega_2)\rangle$ and
 exhibits a non-vanishing imaginary part for broken time-inversion symmetry. 
 Bispectra of continuous quantum measurements have previously been used in the context of quantum transport experiments in nano-electronics \cite{ubbelohdeNATCOMM2012,sifftPRR2021} and for    
 the analysis of non-Gaussian dephasing environments in circuit quantum electrodynamics (cQED) \cite{norrisPRL2016}.
 The cut $S^{(4)}_z(\omega_1,\omega_2,-\omega_1)$ through the fourth-order spectrum (trispectrum) 
  is related to 
$\langle z(\omega_1) z^*(\omega_1) z(\omega_2) z^*(\omega_2) \rangle - \langle z(\omega_2) z^*(\omega_2) \rangle \langle z(\omega_1) z^*(\omega_1) \rangle$ and can be interpreted as an intensity correlation between two frequency contributions to $z(t)$ [exact definition via Eq. (\ref{eq:defPolyspectra})]. 

It is important to note that the second-order spectrum $S^{(2)}$ of a continuous quantum measurement reveals only parts of the information on a measured quantum system.  It is e.g. known, that the spectrum $S^{(2)}$ of telegraph noise of the in- and out-tunneling of an electron to a quantum dot contains only information on the sum of the two tunneling rates. A simultaneous evaluation of $S^{(2)}$ and $S^{(3)}$ is necessary to obtain both tunneling rates separately \cite{sifftPRR2021}. Similarly, the fourth-order spectrum $S^{(4)}$ of a coupled spin-spin system was shown to depend very sensitively on the symmetry of the spin coupling while $S^{(2)}$ does not \cite{hagelePRB2018}. Quantum polyspectra of order $n=3$ and $n=4$ are therefore an indispensable tool for a thorough investigation of quantum systems. The practical evaluation of experimental polyspectra from a finite amount of data using unbiased cumulant estimators and the fast Fourier transformation is described in our Ref. \cite{sifftPRR2021}. Polyspectra of fifth or higher order do - to the best of our knowledge - not appear in literature. The reason is probably a large numerical effort for their calculation and a strongly increasing noise of the measured cumulants for higher orders \cite{schefczikARXIV2019}. Their determination may often yield no meaningful results.

Quantum mechanical expression for $S_z^{(n)}$ for general quantum systems up to order $n=4$ have only recently been derived by our group \cite{hagelePRB2018,hagelePRB2020E}. 
They directly connect the measurable polyspectra of $z(t)$ with properties of the quantum system.
This has paved the way for establishing a new evaluation scheme for transport experiments spanning the full regime between Gaussian and telegraph behavior of the detector output \cite{sifftPRR2021}. The derivation of quantum polyspectra is based on the so-called stochastic master equation (SME) for the system density matrix $\rho(t)$ and the detector output $z(t)$. The SME covers the quantum dynamics of the system including external damping and measurement back-action. The framework of the SME also gives an expression for the time-dependent detector output.  
A continuous monitoring for a measurement operator $A$ yields a 
detector output \cite{tilloyPRA2018,hagelePRB2018}
\begin{equation}
	z(t)  =  \beta^2 {\rm Tr}[\rho(t) (A + A^\dagger)/2]+ \beta \Gamma(t)/2, \label{SME_detector}
\end{equation}
where $\Gamma(t) = dW(t)/dt$
is white Gaussian noise with $\langle \Gamma(t) \Gamma(t') \rangle = \delta(t-t')$ and the differential $dW$ relates to a stochastic Wiener-process. The information on $A$ therefore always comes with Gaussian background noise. The relative background noise reduces for an increasing measurement strength $\beta$.
The system propagates during measurement stochastically via (Ito-calculus)
\begin{eqnarray}\label{eq:sme}
	\mathrm{d}\rho &=& \frac{i}{\hbar}[\rho, H]\,dt  +  \sum_{j}\gamma_j \mathcal{D}[d_j](\rho) \,dt \nonumber \\
	& & + \beta^2 \mathcal{D}[A](\rho) \,dt + \beta \mathcal{S}[A](\rho)\,dW 
\end{eqnarray} 
where the Hamiltonian $H$ governs the coherent evolution of the system matrix $\rho(t)$.
The $d_j$s are jump operators that enter the SME via super operators  
\begin{eqnarray}
	\mathcal{D}[d](\rho) =  d\rho d^\dagger - (d^\dagger d \rho + \rho d^\dagger d)/2.
\end{eqnarray}
that describe damping of the system by the environment ($\mathcal{D}[d_j](\rho) $) and by the measurement process itself ($\beta^2 \mathcal{D}[A](\rho) $). 

The last term in the SME describes via $dW$ a stochastic measurement back-action, where
\begin{equation}
	\mathcal{S}[A](\rho) = A \rho + \rho A^\dagger - \mathrm{Tr}\left[ (A + A^\dagger)\rho \right] \rho
\end{equation}
is non-linear in $\rho$.
The SME has been derived in various forms and varying generality and was rediscovered several times in literature  \cite{barchielliNC1982,barchielliBOOK2009,belavkinConf1987, diosiPLA1988,gagenPRA1993,korotkovPRB1999,korotkovPRB2001,goanPRB2001}.  
An especially intuitive way
of deriving the SME was given by Gross {\it et al.} \cite{Gross2018} and similarly by Atal {\it et al.} \cite{Attal2006, Attal2010}.
They introduce a continuous
sequence of two-level quantum systems (qubits) that
each weakly interact for a short period with the system. They are initially prepared in an equal 
superposition of both states labeled $-1$ and $1$. A projective measurement into the basis states after interaction leads to 
measurement results $-1$ or $1$. Depending on the state of the quantum system the interaction may lead to a slight tendency towards the $-1$ or the $1$ result thus revealing some information on the system. Averaging several measurements in a finite time interval leads to Gaussian noise in the measurement record and some finite offset caused by the measured system. These contributions are reflected in the expression for $z(t)$ by the noise term $\beta \Gamma(t)/2$ and the offset term $\beta^2 {\rm Tr}[\rho(t) (A + A^\dagger)/2]$, respectively. In case of a spin noise measurement, the first term was interpreted as laser shot noise and the second term as a Faraday-rotation signal of the laser polarization that follows the dynamics of the observed spin  \cite{hagelePRB2018}. Moreover, the SME has been shown to correctly describe the quantum system dynamics also in the case of an increasing measurement strength $\beta$. Strong measurements cause the system to quickly collapse into an eigenstate of the measurement operator $A$ which can strongly suppress coherent dynamics and lead to telegraph noise in $z(t)$. This so-called quantum Zeno-effect was discovered by Misra and 
later recovered within a SME treatment \cite{misraJMP1977,korotkovPRB2001,hagelePRB2018}. It came therefore not as a surprise that it was possible to unify the weak and strong measurement regimes of quantum transport into one theory starting from a suitable SME \cite{sifftPRR2021}.

In 2018, three groups were independently able to find quantum mechanical expressions for multi-time moments of $z(t)$ directly from the SME \cite{atalayaPRA2018,tilloyPRA2018,hagelePRB2018}. The expressions are given in terms of the system Liouvillian ${\cal L}[\beta]$ 
\begin{equation}
 {\cal L}[\beta] \rho = \frac{i}{\hbar}[\rho, H]  +  \sum_{j}\gamma_j \mathcal{D}[d_j](\rho) + \beta^2 \mathcal{D}[A](\rho)  
\end{equation}
that covers via 
${\cal L}[\beta] \rho\,dt$ all RHS terms of Eq. (\ref{eq:sme}) that are linear in $\rho$ (with the exception of the stochastic back-action term 
$\beta \mathcal{S}[A](\rho)\,dW$
which is non-linear in $\rho$). Compact expressions are found after further defining the system propagator ${\cal G}(\tau) = e^{{\cal L} \tau}\Theta(\tau)$ with Heaviside-step-function $\Theta(\tau)$, the steady state $\rho_0 =   {\cal G}(\infty) \rho(t)$,
and the measurement super operator ${\cal A} x = (A x + x A^\dagger)/2$  \cite{hagelePRB2018}. 
The general multi-time moments follow then from the SME without any approximation as
\begin{eqnarray}
  \langle z(t_n) &&\cdots z(t_1) \rangle = \nonumber\\
  &&\beta^{2n} {\rm Tr}({\cal A}{\cal G}(t_n - t_{n-1}){\cal A} \cdots {\cal G}( t_{2}-t_1) {\cal A}\rho_0 )
  \label{eq:MomentsK}
\end{eqnarray}
where time order $t_n > t_{n-1} > ... > t_1$ is assumed.
Quantum mechanical expressions for multi-time moments in the form of Eq. (\ref{eq:MomentsK}) have been given in the literature before for several special cases. As early as 1977, Srinivas found a corresponding expression in the context of photon counting probabilities \cite{srinivasJMP1977,zoller1997}.
Bednorz {\it et al.} derive a moment generating functional within a path integral theory  assuming a weak measurement limit and evaluate the functional to arrive at  Eq. (\ref{eq:MomentsK})   [Ref. \onlinecite{bednorzNJP2012}, Eq. (17)]. Wang and Clerk find the same functional as Bednorz via a Keldysh approach and use it to calculate "Keldysh-ordered" moments, cumulants, and spectra of quantum noise up to third order [Ref. \onlinecite{wangPRB2020}, Eq. (3)]. Bednorz and Wang, however, treat the measurement back-action only in the lowest order of $\beta$, while the derivations that use the SME find Eq. (\ref{eq:MomentsK}) without that restriction.

In statistics, cumulants instead of moments are often used for characterizing stochastic processes as cumulants allow for a simple subtraction of additive background noise for all orders \cite{hagelePRB2018}.  We recently derived expressions for quantum mechanical multi-time cumulants of $z(t)$ up to fourth order where the introduction of a modified propagator ${\cal G}'(\tau) = {\cal G}(\tau)-  {\cal G}(\infty)\Theta(\tau)$ and a modified measurement super operator 
 ${\cal A}' x = {\cal A} x
 - {\rm Tr}({\cal A}\rho_0)x$ greatly simplified the notation \cite{hagelePRB2018,hagelePRB2020E}. The quantum polyspectra followed after Fourier transformation: 
\begin{eqnarray}
		S_z^{(2)}(\omega) &=& \beta^4 ( {\rm Tr}[{\cal A}'{\cal G}'(\omega){\cal A}'\rho_0]  + {\rm Tr}[{\cal A}'{\cal G}'(-\omega){\cal A}'\rho_0] ) \nonumber \\
		& & \hspace{-0.3cm} + \beta^2/4 
				\label{eq:S2}
\end{eqnarray}
\begin{eqnarray}		
			S_z^{\rm (3)}(\omega_1,\omega_2,\omega_3 = -\omega_1-\omega_2)  &= & \nonumber \\
	& & \hspace{-40mm}	\beta^6\hspace{-6mm} \sum_{\{k,l,m\} \in \text{prm.} \{1,2,3\}} \hspace{-6mm} {\rm Tr}[{\cal A}'{\cal G}'(\omega_m){\cal A}'{\cal G}'(\omega_m + \omega_l){\cal A}'\rho_0]. \label{eq:S3} 		
\end{eqnarray}
The sum regards all six permutations (prm.) of the indices of the $\omega_j$s \cite{footnote1}. 
The formula for the power spectrum $S^{(2)}_z$ is equivalent to Landau's result \cite{landauBOOKstat} in the absence of damping \cite{hagelePRB2018}. The fourth-order spectrum $S_z^{\rm (4)}$ (see App. \ref{app:QuantumPolyspectra}) exhibits three contributions to the sum.
We emphasize that our expressions for the quantum polyspectra  $S_z^{(3)}$ and $S_z^{(4)}$ have not been known in literature before 2018 and 2020, respectively \cite{hagelePRB2018,hagelePRB2020E}. They are valid for any system that can be described by a Liouvillian ${\cal L}$ in the very general Lindblad-form \cite{lindbladSPRINGER1976} and a measurement procedure described by an operator $A$ that yields information on the observable $(A+A^\dagger)/2$. The spectra exhibit line-broadening due to both environmental and measurement-induced damping including the Zeno-regime of strong measurement. The spectra are free from delta-function contributions because the time-dependent ${\cal G}'(\tau)$ decays exponentially to zero for increasing $\tau$ guaranteeing for a finite ${\cal G}'(\omega)$. 
Quantum polyspectra are straightforwardly evaluated in terms of the eigenvalues and eigenvectors of ${\cal L}$, see Refs. \onlinecite{hagelePRB2018,hagelePRB2020E}.
We consider the general expressions for $S_z^{(3)}$ and $S_z^{(4)}$ a major advancement in the theory of continuous quantum measurements. 

\section{Random-time measurements}
\begin{figure}
	\centering
	\includegraphics[width=8cm]{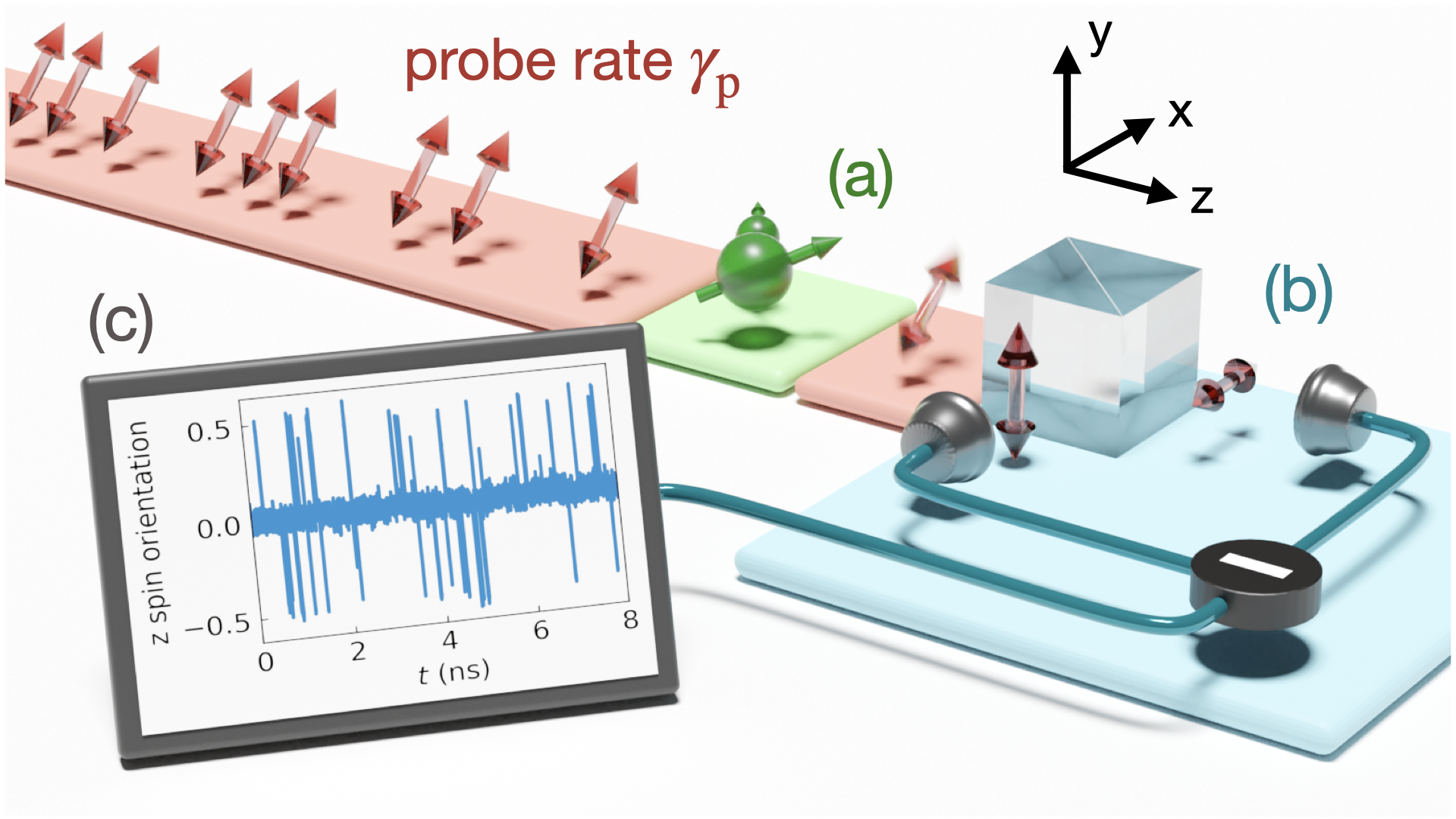}
	\caption{Schematics of random-time measurements. The linearly polarized probe photons (double arrows) arrive at random times in the interaction region (a) where they interact and entangle with the quantum system and may weakly change their orientation (blurred double arrow). 
		After traversing the polarizing beamsplitter a photon with either horizontal or vertical polarization causes an event in the corresponding detector 
		(b) giving rise to a positive or negative peak in the measurement trace (c).	}
	\label{fig:scheme}   
\end{figure}
In this section, we develop a theory of random-time quantum measurements based on the SME and quantum polyspectra up to fourth order. Before, we like to mention that the field of (classical) signal processing knows two similar situations where information on a classical system is gained (or known) only at discrete random times: (i) So-called random-time sampling was developed to determine the power spectrum and the bispectrum (second- and third-order polyspectra) of a stationary stochastic process $x(t)$ via samples $x(t_j)$ taken at random times $t_j$ \cite{shapiroJSIAM1960,pasiniIEEE2005,liiSIGNALPROCESSING1998,benhenniRAM2007}. We are not aware of any theory on fourth-order spectra.
The power spectrum and bispectrum can be recovered even for average sampling rates below the bandwidth associated with the process $x(t)$. Random-time sampling may not be confused with under-sampling where $x(t)$ is sampled at times $t_j = t_0 + j T$, i.e. at a constant rate $1/T$. Spectra determined from under-sampling are known to suffer from replicas of spectral lines with respect to the true spectrum of $x(t)$ and exhibit a small bandwidth limited by $1/T$. Random-time sampling exhibits no replicas and conserves the full bandwidth of $x(t)$. In contrast to random-time quantum measurements the value of $x(t_j)$ is known exactly, while in the quantum case (see below) the dynamic state of the system will only indirectly have an influence on the appearance of click-events in a photo-detector. (ii) Such a behavior is more similar to a so-called Cox process where a classical system determines the probability of a click event \cite{coxJRSSSB1955}. The degree of analogies between classical random-time sampling and random-time quantum measurements is an open question and beyond the scope of the paper.

Fig. \ref{fig:scheme} displays a possible realization of random-time measurements of a quantum system. Probe photons with  $45^\circ$ polarization (red double arrows) interact at random times in region (a) for a short time span with the quantum system. The interaction may slightly change the photon's polarization. 
Subsequently, the photon traverses a polarizing beam splitter with one exit for $0^\circ$ polarized light (x-direction) and another for $90^\circ$ polarized light (y-direction). A single photon will therefore give rise to a click in one of the detectors with near \mbox{50 \%} probability. A positive or negative peak will appear in the measurement record $z(t)$. The exact detection probability will depend on the state of the probed quantum system. Consequently, $z(t)$ will contain some information on the quantum system. The back-action on the system by the probe photon will vary depending on the stochastic outcome of the measurement since system and photon were after their interaction in an entangled state.
Our theory of random-time sampling follows from the stochastic master equation approach by including the dynamics of the system to be measured, the Poisson distributed stream of probe photons, the interaction of a single probe photon with the system, and the readout of the photomultipliers. The measurement operator $A$ will therefore relate to a {\it continuous} measurement of a photon in the photomultipliers and {\it not} to a property of the quantum system as in the usual SME treatment of e.g. a spin noise experiment \cite{hagelePRB2018}. 
Information on the system will nevertheless be obtained as the probe photons interact at {\it random times} with the quantum system. Once ${\cal L}[\beta]$ and $A$ are found for our random-time measurement system, the quantum polyspectra of $z(t)$ follow from the established general equations (\ref{eq:S2}), (\ref{eq:S3}), and (\ref{eq:S4}).

\subsection{The Model Liouvillian}
The overall Hamiltonian $H$ consists of the quantum system (s) and its interaction with the probe photon
\begin{equation}
	H = H_{\rm s} + H_\text{int}, \label{eq:overallHamiltonian}  
\end{equation}
where $H_{\rm s}$ may be any system Hamiltonian and $H_\text{int} = \hbar g s_z a_z$ describes a linear interaction of the system and a probe photon. The factor $g$ corresponds to the interaction strength, $s_z$ relates to some property of the system, and $a_z$ relates to  the photon angular momentum in $z$-direction in the interaction region (a). 
If $s_z$ describes, e.g., an electron spin orientation, the term $H_\text{int}$ models a Faraday rotation of a linearly polarized probe photon in dependence on the  spin orientation $s_z$. A subsequent measurement of the probe photon will thus contain some information on $s_z$. 
The three states for the light mode in our model are $\left| a_{+}\right\rangle$, and $\left| a_{-}\right\rangle$  for the circularly polarized photons and the vacuum state $|a_\text{V}\rangle$ if no photon is present.
The operator $a_z$ thus reads in that basis 
\begin{eqnarray}
	a_z &=&  \mathbb{1}_\text{s} \otimes \left[ \frac{1}{2} \left| a_{+}\right\rangle  \left\langle a_{+} \right| -\frac{1}{2} \left| a_{-}\right\rangle  \left\langle a_{-} \right|    \right]   \otimes \mathbb{1}_\text{b}\nonumber \\
	&=& \mathbb{1}_\text{s} \otimes 
	\begin{pmatrix}
		1/2 & 0&0 \\
		0& -1/2&0\\
		0&0&0
	\end{pmatrix}
	 \otimes \mathbb{1}_\text{b},
\end{eqnarray}
where we recognize the Pauli spin operator for the $z$-direction. The unit-operators $\mathbb{1}_\text{s}$ and $\mathbb{1}_\text{b}$ belong to the Hilbert-spaces that describe the quantum system (s) and the detector region (b), respectively.

The stream of polarized photons entering the interaction area (a) is modeled within the SME analogously to the incoherent in-tunneling of electrons onto a quantum dot \cite{sifftPRR2021}. The quantum state of a $45^\circ$ polarized photon is given by $\left| a_{45}\right\rangle = \frac{1}{\sqrt{2}}( \left|a_{+}\right\rangle  +  i \left| a_{-}\right\rangle)$. 
 The jump operator 
\begin{eqnarray}
	d_\text{a} = \mathbb{1}_\text{s} \otimes \left| a_{45}\right\rangle  \left\langle a_\text{V} \right|  \otimes \mathbb{1}_\text{b},
\end{eqnarray} 
creates a $45^\circ$ polarized photon in the interaction area. Probe photons appearing at an average rate $\gamma_\text{p}$ give rise to a term $\gamma_\text{p}{\cal D}[d_\text{a}] (\rho)$ at the RHS of the SME. 

The interaction region is then emptied at a rate $\gamma_\text{out}$ via again an incoherent tunneling process into the detector region (b).
For $\gamma_\text{out} \gg \gamma_\text{p}$  the appearance of photons in the interaction region is Poisson distributed.
The incoherent tunneling must conserve the polarization state of the photon. This is accomplished by   
\begin{eqnarray}
d_\text{ab} &=& \left( \mathbb{1}_\text{s} \otimes  \left| a_\text{V} \right\rangle  \left\langle a_{+} \right| \otimes \left| b_{+} \right\rangle  \left\langle b_\text{V} \right| \right) \nonumber \\
&+& \left( \mathbb{1}_\text{s} \otimes  \left| a_\text{V} \right\rangle  \left\langle a_{-} \right| \otimes  \left| b_{-} \right\rangle  \left\langle b_\text{V} \right| \right) \label{eq:coherentJump}
\end{eqnarray}
and $\gamma_\text{out}{\cal D}[d_\text{ab}] (\rho)$ on the RHS of the SME. We emphasize that $ d_\text{ab}$ does not change if instead of the circular polarized basis states the linear or any other basis states are used. 
The polarizing beam splitter will split the beam of photons in $0^\circ$ (x-direction) and $90^\circ$ (y-direction) polarized photons described by
$\left| b_{0}\right\rangle = \frac{1}{\sqrt{2}}( \left|b_{+}\right\rangle  +   \left| b_{-}\right\rangle)$ and $\left| b_{90}\right\rangle = \frac{1}{\sqrt{2}}( \left|b_{+}\right\rangle  - \left| b_{-}\right\rangle) $, respectively. 
The detection of the two polarization states and the no-photon state is modeled by projective measurements via the operators $ | b_{0}\rangle \langle b_{0} |$,  $| b_{90}\rangle \langle b_{90}| $, and $ | b_\text{V}\rangle \langle b_\text{V} |$ which relate to output values $1/2$, $-1/2$, and $0$, respectively.  
The measurement operator $A$ thus reads 
\begin{equation}
 A = \mathbb{1}_\text{s} \otimes \mathbb{1}_{\rm a} \otimes \left[  | b_{0}\rangle \langle b_{0} | -  | b_{90}\rangle \langle b_{90} | \right]/2 
 \label{eq:measurementOperator}
\end{equation}
which will result in a positive or negative peak in the measurement trace $z(t)$ depending on which polarization is detected.

The disappearance of the photon from the detector region is obtained by jump operators that model the transitions from the photon states to the vacuum state:
\begin{eqnarray}
d_{{\rm be}, 0 / 90  } = \mathbb{1}_\text{s} \otimes \mathbb{1}_{\rm a} \otimes 
\left| b_\text{V}\right\rangle  \left\langle b_{0/90}   \right|.
 \end{eqnarray} 
The sum of super operators $\gamma_{\rm det} (\mathcal{D}[d_{{\rm be}, 0 }](\rho)  + \mathcal{D}[d_{{\rm be}, 90 }](\rho))$ 
 appears on the RHS of the SME for photons that leave the detector at rate $\gamma_{\rm det}$. Again, different choices of the basis states leave the sum unchanged.

 The overall Liouvillian is given by
 \begin{eqnarray}
{\cal L}[\beta](\rho) &=& \frac{i}{\hbar}[\rho, H] +  \gamma_\text{p} \mathcal{D}[d_\text{a}](\rho)
+  
\gamma_{\rm out} \mathcal{D}[d_\text{ab}](\rho) \nonumber \\
& &
\hspace{-1.5cm} + \,\gamma_{\rm det} (\mathcal{D}[d_{{\rm be}, 0 }](\rho)  + \mathcal{D}[d_{{\rm be}, 90 }](\rho))  + \beta^2 \mathcal{D}[A](\rho).
\end{eqnarray} 
The first term on the RHS describes the coherent system dynamics and interaction with the probe photon. All other terms model {\it random-time measurements} via the {\it continuous} detection of stochastically arriving probe photons.
 The detector output is given by $z(t) = \beta^2{\rm Tr}[(A \rho + \rho A^\dagger)/2] + \beta \Gamma(t)/2$ and 
 the SME
\begin{equation}
\mathrm{d}\rho = \mathcal{L}[\beta] (\rho)\,dt + \beta \mathcal{S}[A](\rho)\,dW \label{eq:SME_Liou}
\end{equation}	 
describes the dynamics of the combined system.
\subsection{Discussion of Model}

{\it Stochastic interaction time:} The disappearance of probe photons from the interaction region is modeled via incoherent tunneling at the rate $\gamma_{\rm out}$. This implies that the times of interaction vary between the probe events exhibiting an exponential distribution. We argue that a distribution of interaction times in contrast to a fixed interaction time naturally appears in many quantum systems. In case of a semiconductor spin noise experiment, the probe photon enters the semiconductor sample whose surfaces act as semi-transparent mirrors. Consequently, the probe photon finds itself in a lossy optical cavity and will leave the cavity statistically leading to an exponential distribution of interaction times.

The modeling of an (at least approximately) constant interaction time within a Liouvillian approach appears to be quite challenging. One might think of a probe wave packet that traverses a broad interaction region with a constant speed.  The Hilbert space of the interaction region needed to be complemented by a large number of position states that describe the appearance of the probe photon at different sites of the interaction region. 
A similar problem was discussed by Peres. He showed that a quantum clock requires a large number of states to be used for accurate timing of certain events (cmp. Sec. V in \cite{peresAJP1980}).  
Our model requires only one site for the probe photon in the interaction region and is therefore much simpler.  It nevertheless describes the physics of a semiconductor spin noise experiment correctly (regarding our argument above).

We like to mention that Gross {\it et al.} have treated the interaction of a quantum system (see Section 7 in \cite{Gross2018}) and its random-time probe by a fixed interaction time without a reference to a specific quantum mechanical model that would realize such a constant interaction time. They give expressions for the corresponding measurement operators (Kraus operators), and derive an effective stochastic master equation for the average behavior of the quantum system. They, however, neither give an expression for the detector output [our $z(t)$] nor attempt to find expressions for its polyspectra.

{\it Measurement process:} 
Similarly to the stochastic interaction time also the dwell time of the probe photon in the detector region is stochastic. This leads to a realistic detector behavior with a finite response time and varying peak area (see Fig. \ref{trace}). A completely alternative treatment of random-time measurements via Poisson processes in the SME (a $dN$ instead of $dW$ appears in the equations) \cite{jacobsCP2006}
would correspond to an unrealistic instant response of the detector, peaks of constant area, and no Gaussian background noise. 

Moreover, the question may arise, why the measurement was modeled via a projective operator $| b_{0}\rangle \langle b_{0} |$ \mbox{[Eq. (\ref{eq:measurementOperator})]} and not via an annihilation operator \mbox{$c = | b_\text{V}\rangle \langle b_{0} |$} which at the same time would empty the detector region after detection. The problem occurs that $z(t)$ would relate to ${\rm Tr}[(c + c^\dagger)\rho(t)/2]$, [see Eq. (\ref{SME_detector})], which unlike ${\rm Tr}[| b_{0}\rangle \langle b_{0} |\rho(t)]$ is not proportional to the probability of finding the probe photon in the $0^\circ$ polarized state.

%\subsection{Random-time measurements of a single spin}
\subsection{Spin dynamics under single-photon probing}
 \begin{figure}[t]
	\centering
	\includegraphics[width=8cm]{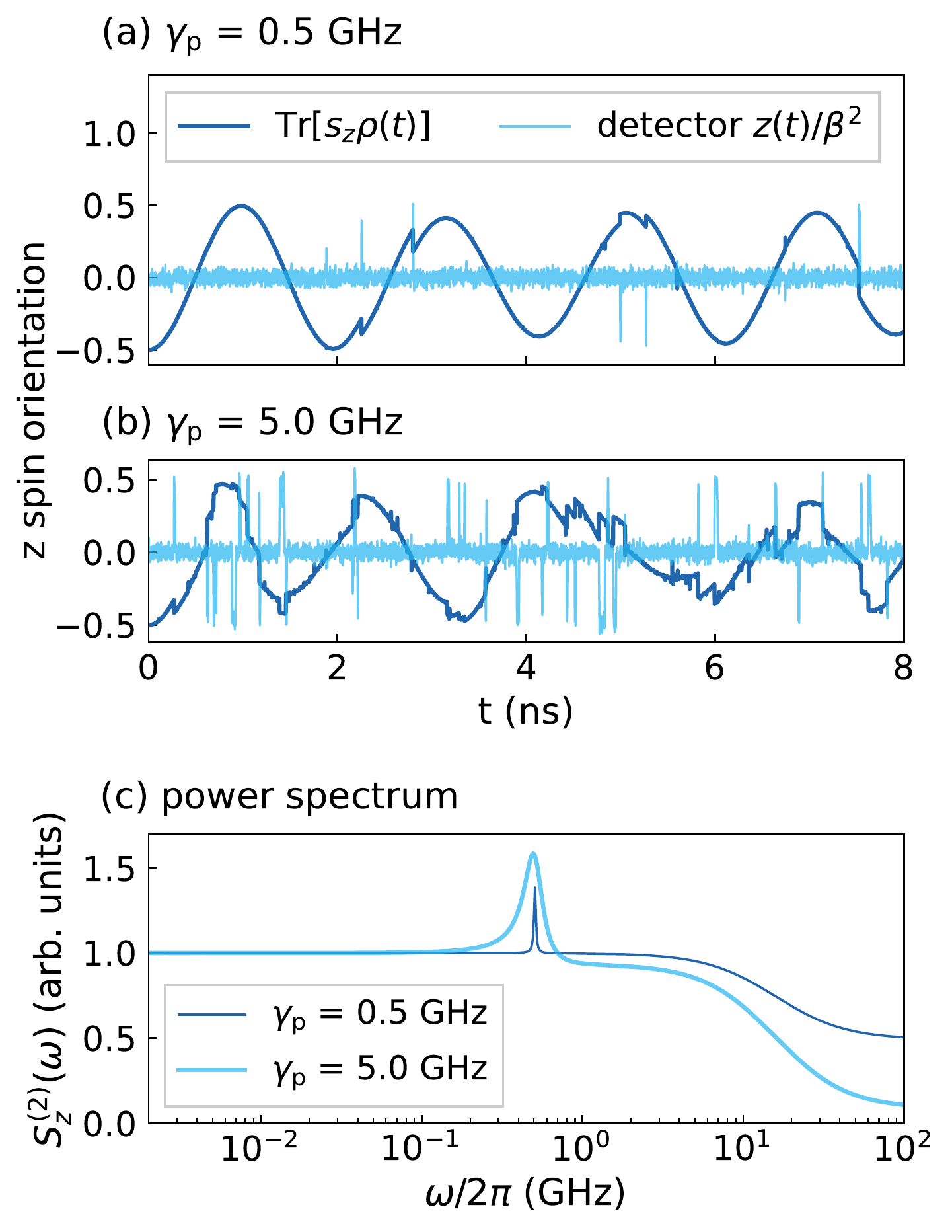}
	\caption{Random-time measurement of the $z$-direction of a single precessing spin via the Faraday effect by a stream of single photons. (a) Detector output shows single click events (thin line) at low sampling rates while the system exhibits an almost coherently precessing spin (bold line).
	(b) The spin precession shows stronger perturbations at elevated average sampling rates. (c) Corresponding power spectra of the detector output for low and elevated average sampling rates.  }
	%Simulated sample traces for a low photon rate (a) and a high photon rate (b) of the detector output $z(t)$ (light blue). The evolution of the spin (black) is disturbed by the measurement backaction in case of a high photon rate. In (c) a comparison between the power spectra of the detector output for the two cases above.}
	\label{trace}   
\end{figure}
Next, we calculate as an illustrative example the detector output and second-order spectrum of a random-time measurement on a single electron spin. The spin is precessing in an external magnetic field $B$ parallel to the $x$-direction with the system Hamiltonian given by
\begin{eqnarray}
H_\text{s} = \hbar \omega_L s_x,
\end{eqnarray}
where 
$\omega_\text{L}$ 
 is the Larmor frequency. 

The $z$-direction of the electron spin is probed via the Faraday effect by a stream of single photons $45^\circ$ polarized in the $xy$-plane (Fig. \ref{fig:scheme}). 
After interacting with the system (s) via $ \hbar g s_z a_z$ [see Eq. (\ref{eq:overallHamiltonian})], the photon polarization axis is slightly rotated in the $xy$-plane depending on the $z$-spin orientation. A measurement of the photon’s polarization rotation thus gives access to the electron spin operator $s_z$.
 The photon is  measured in the $0^\circ$ (x-direction) and $90^\circ$ (y-direction) polarization directions. The detector output $z(t)$ will exhibit either a strong positive or a strong negative peak. Without interaction their probability of appearance is 50~\% each. The probabilities slightly change if the system alters the photon's polarization state in the interaction region. The measurement result, therefore, contains some (but not full) information about the electron's $z$-spin orientation at the time of interaction. 
\begin{figure}[t]
	\centering
	\includegraphics[width=7.5cm]{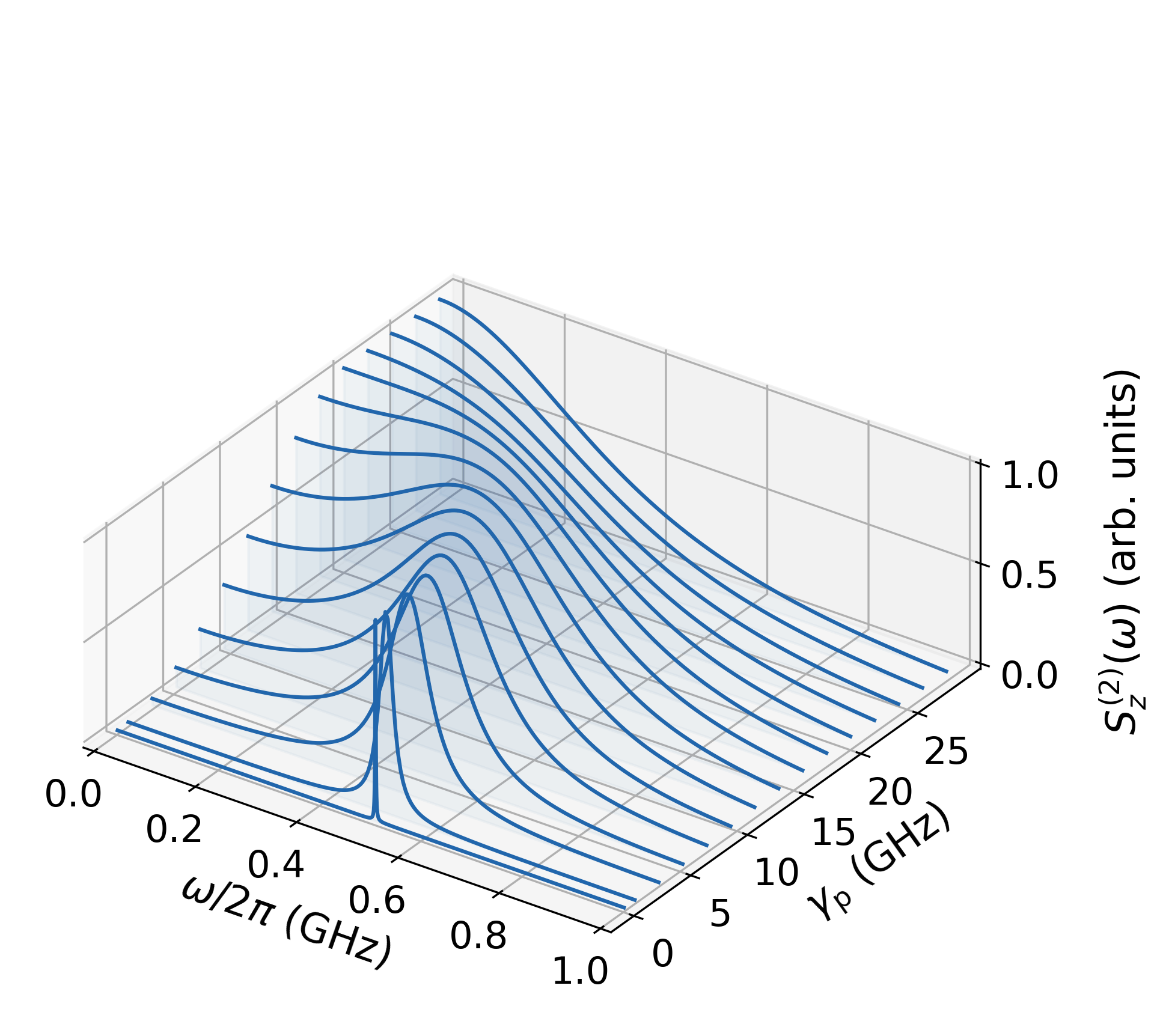}
	\caption{Power spectra of the detector output for increasing average sampling rates $\gamma_\text{p}$ of a precessing electron spin show a clear Zeno transition
	with suppression of precession dynamics for highest rates.}
	\label{fig:zeno}   
\end{figure}

Figure \ref{trace}(a) shows the detector output $z(t)$ found from numerical integration of Eq. (\ref{eq:SME_Liou}).
A sampling rate $\gamma_\text{p} = 0.5$~GHz was chosen which corresponds to a probe laser power of $0.12$~nW at a wavelength of $800$~nm. The tunnel-rate \mbox{$\gamma_\text{out} =  100$~GHz} and the detector-rate \mbox{$\gamma_\text{det} = 100$~GHz} correspond to a detector with roughly $10$~ps temporal resolution as can be realized in experiments with avalanche photodiodes.  The interaction strength $g = 100$~GHz is relatively high and can be obtained via a tight laser focus within a resonant optical micro-cavity \cite{PoltavtsevPRB2014}.
The measurement strength $\beta^2 = 10^4$~GHz ensures a collapse of the photon wavefunction well within the time the photon spends in the detector region.  
The measurement trace $z(t)$  exhibits clear  peaks, both positive and negative, related to photon detection events on an otherwise Gaussian background noise $\beta \Gamma(t)/2$ [cmp. Eq. (\ref{SME_detector})].  The Gaussian background disappears only in the unrealistic limit of ultra-strong measurement of the photon. Any detector circuit for an avalanche photodiode or photo-multiplier exhibits background noise. Appearance of background noise is therefore a desired feature of a realistic detector model. 
 The peaks vary in height and width closely resembling actual current traces of photomultiplier tubes \cite{foord1969use}. 
 The expectation value $\text{Tr}(s_z \rho(t))$ of the spin $z$-component displays a coherent precession dynamics that is only 
 weakly disturbed at the times of a photon detection event [Fig. \ref{trace}(a)]. 
 At an increased average rate $\gamma_\text{p} = 5.0$~GHz of the incoming probe system, the precession dynamics is clearly distorted
  [Fig. \ref{trace}(b)]. 
  Since in quantum mechanics $\text{Tr}(s_z \rho(t))$ can not be directly observed in an experiment, information on the system's dynamics must be deduced from the measurement trace $z(t)$. 
  
Measurement traces which exhibit click events are known from photon counting experiments where usually some electronics searches for peaks in $z(t)$ using suitable threshold criteria. Counting rates or counting statistics follow then from the time-intervals between peaks. In contrast, we will consider here spectra and polyspectra of the raw $z(t)$ without the need for artificial threshold criteria (numerics of experimental polyspectra, see App. B of Ref. \onlinecite{sifftPRR2021}). This approach therefore works without any problems even in cases where Gaussian background noise becomes problematic for identifying single peaks in $z(t)$. 
Here, we obtain the analytic power spectrum $S^\text{(2)}_z (\omega)$ from evaluating Eq. (\ref{eq:S2}) which depends only on the Liouvillian ${\cal L}[\beta]$, the measurement operator $A$, and on the measurement strength $\beta$ \cite{sifftPRR2021}.

Fig. \ref{trace}(c) shows the spectrum for average sampling rate $\gamma_\text{p} = 0.5$~GHz. Three contributions can be distinguished: (i) A constant white noise background due to Gaussian detector noise $\beta \Gamma(t)/2$ clearly visible at \mbox{$\omega/2 \pi \approx$ \SI{100}{GHz}}, (ii) a broad Lorentzian peak centered at \SI{0}{Hz} with a cutoff frequency at around $\SI{16}{GHz}$ which corresponds to the finite lifetime of the probe system in the detector region, (iii) a narrow Lorentzian peak at the Larmor frequency $\omega_\text{L}$ originating from the system dynamics.
  For a higher probe rate $\gamma_\text{p} = 5.0$~GHz,
 the Larmor-peak exhibits a clear broadening and small shift to lower frequencies as expected for a precessing spin subject to increased measurement induced damping. 
Fig. \ref{fig:zeno} shows the power spectra for  increasing measurement rates $\gamma_\text{p}$ where the background has been subtracted using power spectra for the case of no probe-interaction with the system, $H_{\rm int} = 0$.
At high rates the 
spectrum broadens and shifts to zero frequencies as the frequent measurements suppress all coherent dynamics. This behavior is known as quantum Zeno effect \cite{misraJMP1977}.
Previously, Korotkov studied the Zeno transition using a continuous measurement approach. He found the same spectral features as in our case of random sampling \cite{korotkovPRB2001}.  Zeno physics can in principle also lead to the suppression of the coupling to the environment and consequently to the suppression of decoherence.  Such behavior has been discussed by Gordon {\it et al.} \cite{gordonJPB2007a,gordonJPB2007b} investigating a stochastic modulation of the system and an environment with finite correlation time. In contrast, the Lindblad type Liouvillian of our theory implies a treatment of the environment induced decoherence in Markov approximation where decoherence cannot be suppressed by frequent measurements \cite{KofmanPRA2001}.  

Our numerics for $\gamma_\text{p}$ down to $0.05$~GHz suggests that the width of the spectrum scales for lower rates linearly with $\gamma_\text{p}$ allowing for the detection of a fully coherent oscillation in the limit $\gamma_\text{p}
\rightarrow 0$. This is in agreement with Ruskov's early theory for a two-level system \cite{ruskovPhysRevB2003}. Similarly, we found for general systems a sharpening of  spectral features also in the third-order polyspectra (not shown).
%The disturbance of the system is finite at any single probing event (see discrete distortion in $z(t)$ [Fig. 2(a)]) and – unlike in the continuous case – never approaches zero.
 Nevertheless, an increase in interaction strength $g$ leads to a stronger disturbance of the system at any single probing event (see discrete distortion in $z(t)$ [Fig. 2(a)]). This changes the height of spectral features even for $\gamma_\text{p} \rightarrow 0$ revealing quantum back-action.

%An analytical proof for this behavior is, however, still missing.
\begin{figure*}
	\centering
	\includegraphics[width=15.7cm]{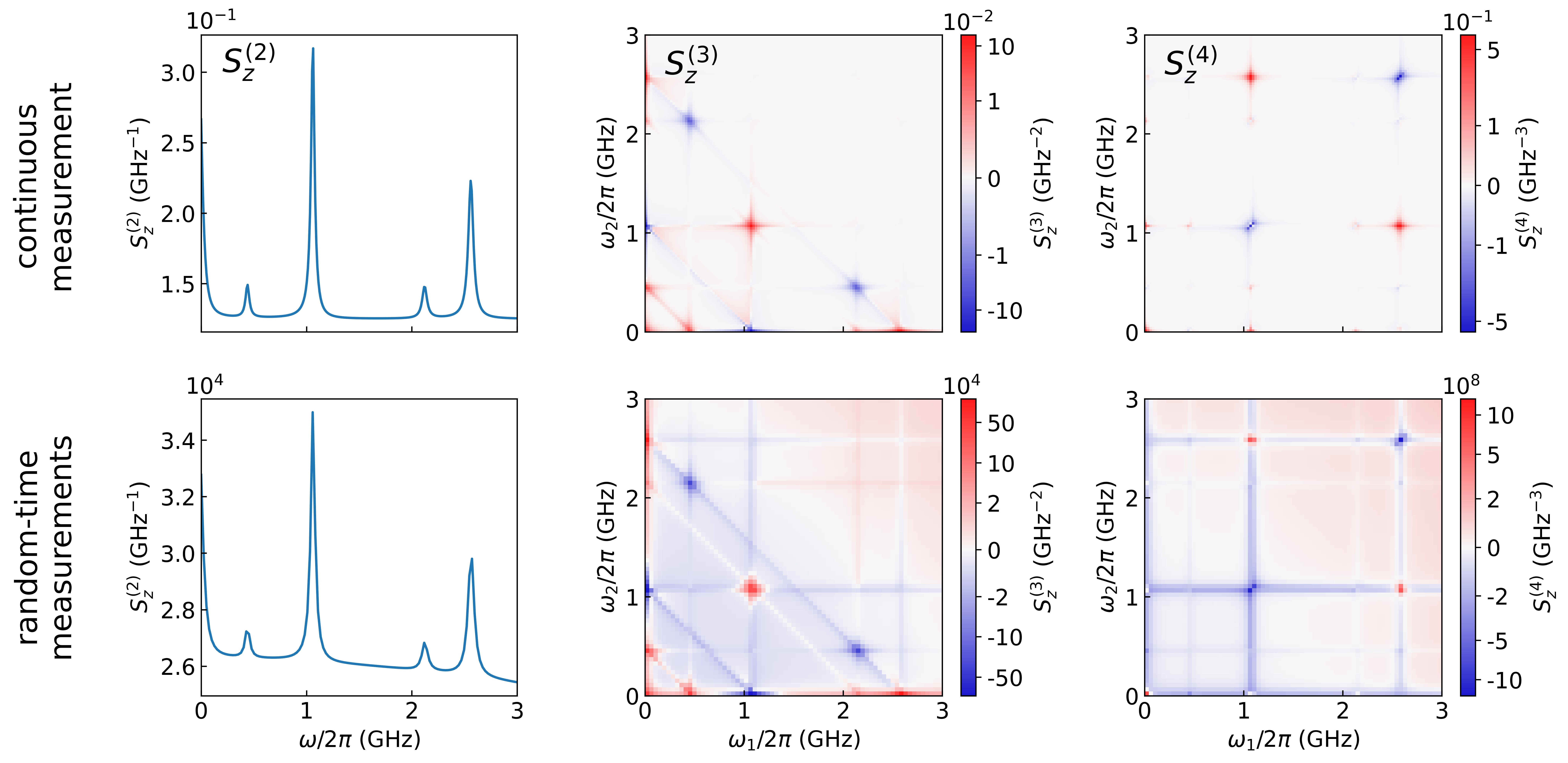}
	\caption{Power spectrum $S^\text{(2)}_z (\omega)$, bispectrum $S^\text{(3)}_z (\omega_1,\omega_2)$, and trispectrum $S^\text{(4)}_z (\omega_1,\omega_2)$ of the $z$-component of one spin in a coupled spin system for continuous quantum measurements in comparison with random-time measurements.  An average background of $-3.88\times 10^4$~GHz$^{-2}$ and $-6.60\times 10^8$~GHz$^{-3}$ was subtracted in the random-time bi- and trispectrum, respectively. The spectra closely resemble each other except for the background and an additional stripe structure. The color bar is scaled via the arsinh-function.}
	\label{fig:bispectrum}   
\end{figure*}

\subsection{Polyspectra of a two-spin system}
Next, we calculate second-, third-, and fourth-order quantum polyspectra (power spectrum, bispectrum, and trispectrum) of the random-time measurement of a coupled spin-spin system in a tilted magnetic field \cite{hagelePRB2018}. Higher-order polyspectra had previously been shown to yield important information that is not contained in the usual second-order spectra \cite{sifftPRR2021}. We compare the spectra with traditional spectra for continuous sampling and find the same structure with a number of additional stripes in the case of third- and fourth-order spectra. This establishes the central result of our paper: the possibility to characterize a quantum system by random-time quantum measurements without any loss of information in comparison to the well established continuous quantum measurements.   

The quantum system is defined by
\begin{eqnarray}
H_\text{s}  & = &  \hbar \omega^{(1)}_\text{L} (\sin(\varphi) s^{(1)}_x + \cos(\varphi) s^{(1)}_z) \nonumber \\
 & & +  \hbar \omega^{(2)}_\text{L} (\sin(\varphi) s^{(2)}_x + \cos(\varphi) s^{(2)}_z) \nonumber \\
  & & +  \hbar g_\text{c} [ s^{(1)}_x s^{(2)}_x  +  s^{(1)}_y s^{(2)}_y + s^{(1)}_z s^{(2)}_z ],
\end{eqnarray}
with the precession frequencies $\omega^{(1)}_\text{L}/2 \pi = 1.5$~GHz, $\omega^{(2)}_\text{L} / 2 \pi= 0.0$~GHz, and an isotropic spin-spin interaction with the coupling strength
\mbox{$g_\text{c} / 2 \pi = 1.5$~GHz}. The tilt-angle of the magnetic field is chosen to be  \mbox{$\varphi = \pi/6$}. 
The $z$-component of the first spin is coupled to the probe-photons via
$
  H_\text{int}= \hbar g s^{(1)}_z a_z,
$ 
where $g = 50$~GHz. The probe photon enters the interaction region at rate \mbox{$\gamma_\text{p} = 5$~GHz} and transfers to the detector region
 at rate $\gamma_\text{out} = 100$~GHz. 
 The measurement operator  $A$ for the polarization bridge  is similar to Eq. (\ref{eq:measurementOperator}) with a measurement strength $\beta^2 = 10^4$~GHz.  The photon leaves the detector at the rate $\gamma_\text{det} = 100$~GHz.
An additional spin relaxation term $\frac{\gamma_{\rm s}}{2} \mathcal{D}[d_\text{s}](\rho)$ with
$
d_{\rm s} = \left| s^{(1)}_{z,\downarrow}\right\rangle  \left\langle s^{(1)}_{z,\uparrow}  \right| \otimes \mathbb{1}_{{\rm s}^{(2)}} \otimes
\mathbb{1}_\text{a} \otimes \mathbb{1}_{\rm b}
$
is introduced which drives the first spin towards the $-z$ direction at a slow rate \mbox{$\gamma_{\rm s} = 0.05$~GHz}. 
Consequently, the $z$-spin orientation is negative in equilibrium implying  ${\rm Tr}( s^{(1)}_z \rho_0) < 0$ and $\langle z(t) \rangle < 0$. 
The spectrum $S^\text{(2)}_z (\omega)$ of the random-time measurement trace in \mbox{Figure \ref{fig:bispectrum}} shows several peaks corresponding to quantum beats between different energy eigenstates of the system. As for the single-spin example, a broad Lorentzian background (spectrum of single clicks) is visible in the case of random-time sampling.
% Should the detector lifetime be known the Lorentzian background could be subtracted. (See analytic form of telegraph spectra.) 
The corresponding spectrum for continuous measurements (calculated via $H_{\rm int} = 0$ , $A = s^{(1)}_z$,  $\beta^2 = 0.5$~GHz) shows only its typical flat white noise background \cite{hagelePRB2018}. 

Additional spectral contributions appear also in the higher-order spectra. The third-order spectrum, Eq. (\ref{eq:S3}), exhibits a large negative background (subtracted in Fig. \ref{fig:bispectrum}) and several sharp peaks corresponding to the ones seen in the bispectrum $S^{(3)}$ of the usual continuous measurement. In addition, straight lines that extend to higher frequencies are visible in the random-time case.
Benhenni and Rachdi show in their classical treatment of bispectra from random-time sampling that 
the asymptotic behavior of the additional strip structure to $S^{(3)}$ allows for a separation from the desired peak structure \cite{benhenniRAM2007}.  After a first attempt we believe that separation of such contributions is also possible in the quantum case. As observed in numerical studies, the mean value of $z(t)$ correlates with the appearance of the stripe structure. A third-order spectrum without background and stripes was found in our example only for $\langle z(t) \rangle = 0$ when $\gamma_{\rm s} = 0$. In that regime the average number of positive and negative clicks in the measurement trace is the same.

Similarly to the third-order spectrum, the fourth-order spectrum $S^{(4)}(\omega_1,\omega_2,-\omega_1)$ exhibits peaked contributions that correspond to the usual fourth-order spectrum. In addition, striped contributions appear that extend to higher frequencies on an overall 
 large offset due to the Poisson noise contribution (subtracted in Fig. \ref{fig:bispectrum}). Thus, the fourth-order spectra in the random-time measurement regime contain the same information as in the continuous regime \cite{hagelePRB2018}.  As far as we know, the fourth-order spectrum of classical random-time sampling has not yet been addressed in literature. 
%The parameters of a quantum system can in principle by obtained by fitting measured polyspectra to quantum polyspectra calculated from the Liouvillian. It is important to note that the power spectrum alone does not provide enough information to determine system parameters such as the coupling strength, $g_\text{c}$.
Our approach allows also for modeling detectors with a finite response time and finite detection efficiency (determined via $\gamma_\text{out}$ and $\gamma_\text{det}$). We find qualitatively the same peak structure in all polyspectra for different detector parameters albeit with varying strengths of the background offset and the stripe features.

%Although it may seem that system dynamics are not directly accessible in the case of random sampling, polyspectra do provide a key to system characterization. Any artifacts (strips or background) that arise from an actual laboratory measurement in the random sampling regime also arise in the analytical calculation of the spectra. This allows the determination of the system parameters based on the fit between measured and simulated polyspectra.

\section{Discussion}
We introduced a very general framework for treating random-time quantum measurements. This enables a thorough characterization of quantum systems, even at low average sampling rates by comparing the theoretical and measured higher-order spectra \cite{sifftPRR2021}. Any photon-counting experiment fulfilling steady-state conditions can be treated within that framework. We correctly describe quantum back-action and treat environmental damping within Markov approximation like with any Lindblad master equation. Many real-world features of experiments like finite temporal resolution of detectors, background noise, or photon loss ($\gamma_{\rm det} \gg \beta^2$) can be modeled within that framework. 

Therefore, the solution to several open problems in measurement theory becomes apparent. This includes the problem of determining the one- and off-switching rates of fluorophores in microscopy at low photon rates. The problem appears when the typical transients of telegraph noise vanish in the photon noise and switching events can no longer be identified \cite{lippitzCPC2005}. Since our work shows that random-time sampling conserves all information even at low photon rates, we will be able to characterize switching dynamics also in the - previously inaccessible - low sampling rate regime.   Moreover, random-time sampling offers a solution to the
problem of evaluating spin noise spectra at very low probe laser intensities where only single-photon events can be detected.  Random sampling is therefore an attractive alternative to amplification schemes for the probe laser via heterodyning that had been used to circumvent the problem \cite{cronenbergerRSI2016,sterinPRAPL2018,kamenskiiPRB2020}.  
Furthermore, random-time measurements are a viable and more versatile alternative to the recently introduced high-resolution spectroscopy via sequential weak measurements which exhibit unwanted spectral replicas \cite{pfenderNATCOMM2019}. Another promising application lies in today's experiments of circuit quantum electro-dynamics which offer a great control of probe events with desired timing and interaction strength \cite{ficheuxNATCOMM2018,minevNATURE2019}. Especially, the investigation of non-Gaussian environmental noise in cQED may benefit from measuring polyspectra of a detector qubit via random-time measurements allowing for ultra-weak back-action \cite{norrisPRL2016}.

Applications are in reach as real-time spectrometers for $S^{(2)}$ with GHz-bandwidths are available from several vendors. Even, higher-order spectrometers for $S^{(3)}$ and $S^{(4)}$ have been realized by Balk {\it et al.}  and Starosielec {\it et al.}, respectively \cite{balkPRX2018,starosielecRSI2010}. Our framework can therefore be applied without any impediments.

\begin{acknowledgments}
	We acknowledge financial support by the German Science Foundation (DFG) under Project No. 341960391.
\end{acknowledgments}

\begin{widetext}

\appendix

\section{Fourth-order quantum polyspectrum}
\label{app:QuantumPolyspectra}
The fourth-order polyspectrum of the detector output $z(t)$  of the continuously monitored quantum system in the steady state follows from the SME
without any approximations as (second- and third-order spectra see main text) \cite{footnote1}
	\begin{eqnarray}	
		S_z^{\rm (4)}(\omega_1,\omega_2,\omega_3,\omega_4 = -\omega_1-\omega_2-\omega_3) & = & \hspace{4mm}\beta^8 \hspace{-8mm} \sum_{\{k,l,m,n\} \in \text{prm.} \{1,2,3,4\}}
		\hspace{-8mm} \left[ {\rm Tr}[{\cal A}'{\cal G}'(\omega_n){\cal A}' {\cal G}'(\omega_m + \omega_n){\cal A}'{\cal G}'(\omega_l + \omega_m + \omega_n){\cal A}'\rho_0] \right.  \label{eq:S4} \\  \nonumber
		&-& \frac{1}{2 \pi}\int{\rm Tr}[{\cal A}'{\cal G}'(\omega_n) {\cal G}'(\omega_m + \omega_n - \omega){\cal A}'\rho_0]{\rm Tr}[{\cal A}'{\cal G}'(\omega) {\cal G}'(\omega_l +\omega_m + \omega_n){\cal A}'\rho_0]\textrm{d}\omega \\ \nonumber
		&-& \left. \frac{1}{2 \pi}\int{\rm Tr}[{\cal A}'{\cal G}'(\omega_n) {\cal G}'(\omega_l +\omega_m + \omega_n){\cal G}'(\omega_m + \omega_n - \omega){\cal A}'\rho_0]{\rm Tr}[{\cal A}'{\cal G}'(\omega) {\cal A}'\rho_0]\textrm{d}\omega\right]. \label{eq:S4}
	\end{eqnarray}
	The derivation via multi-time cumulants of $z(t)$ and an efficient method for its numerical evaluation are given in Refs. \onlinecite{hagelePRB2018,hagelePRB2020E}. Numerics is based on the QuTiP and ArrayFire software libraries \cite{JOHANSSON20131234,Yalamanchili2015}
\end{widetext}

%\newpage
%


\begin{thebibliography}{68}
\bibitem [{foo()}]{footnote1}%
	\BibitemOpen
	\href@noop {} {\bibinfo {title} {{Please note, that $k$ is not required on
				the right-hand side.}}}\BibitemShut {Stop}%
\end{thebibliography}
\end{document}